\documentclass[letterpaper]{dae-handout}
\graphicspath{{./}{./FIGURES/}{./figures/}{../Figures/}{../../Figures/}{./images/}{./graphics/}}

\newif\ifarxiv
\arxivtrue

\newcommand{\trversion}{v0.4 DRAFT}
\newcommand{\trfilename}{borrill-circumventing-flp-2026.tex}
\newcommand{\trdate}{2026-02-23}
\newcommand{\trshorttitle}{Circumventing FLP with OAE}
\newcommand{\trauthor}{Paul Borrill}
\newcommand{\traffiliation}{D\AE D\AE LUS}

\usepackage[T1]{fontenc}
\usepackage[utf8]{inputenc}
\usepackage[bitstream-charter]{mathdesign}   % Leibniz Bridge standard font (has bold small caps)
\usepackage{microtype}
\usepackage{amssymb}
\usepackage{amsmath}
\usepackage{amsthm}
\usepackage{booktabs}
\usepackage{enumitem}
\usepackage{fancyhdr}
\usepackage{titletoc}
\usepackage{etoc}
\usepackage{lastpage}
\usepackage{xcolor}
\usepackage{tikz}
\usepackage{eso-pic}
\usepackage{url}
\usepackage{morefloats}   % increase float limit for Tufte margin notes
\usepackage{placeins}     % \FloatBarrier
\usetikzlibrary{positioning,shapes.geometric}

\setcitestyle{numbers,square}

\newtheorem{theorem}{Theorem}

\definecolor{linkblue}{RGB}{30,80,180}
\definecolor{linkgreen}{RGB}{30,120,60}
\definecolor{linkred}{RGB}{140,30,30}

\hypersetup{
  colorlinks=true,
  urlcolor=linkblue,
  linkcolor=linkgreen,
  citecolor=linkgreen,
}

\makeatletter
\ifarxiv
  \newcommand{\sidecite}[3][]{\citep{#2}}
\else
  \newcommand{\sidecite}[3][]{%
    \def\sc@temp{#1}%
    \citep{#2}%
    \ifx\sc@temp\empty
      \marginnote{\scriptsize\citep{#2}\enspace #3}%
    \else
      \marginnote[#1]{\scriptsize\citep{#2}\enspace #3}%
    \fi
  }
\fi
\makeatother

\tikzset{
  badge/.style={
    circle,
    draw=#1,
    fill=#1!10,
    line width=0.5pt,
    minimum size=0.45in,
    font=\tiny\sffamily,
    align=center,
    text=#1!80!black
  }
}

\newcommand{\placebadges}{%
  \ifarxiv\else
  \AddToShipoutPictureBG*{%
    \AtPageUpperLeft{%
      \raisebox{-0.5in}{\hspace{\dimexpr\paperwidth-0.7in\relax}%
        \begin{tikzpicture}[overlay]
          \node[inner sep=0pt] (logo)
            {\includegraphics[height=0.855in]{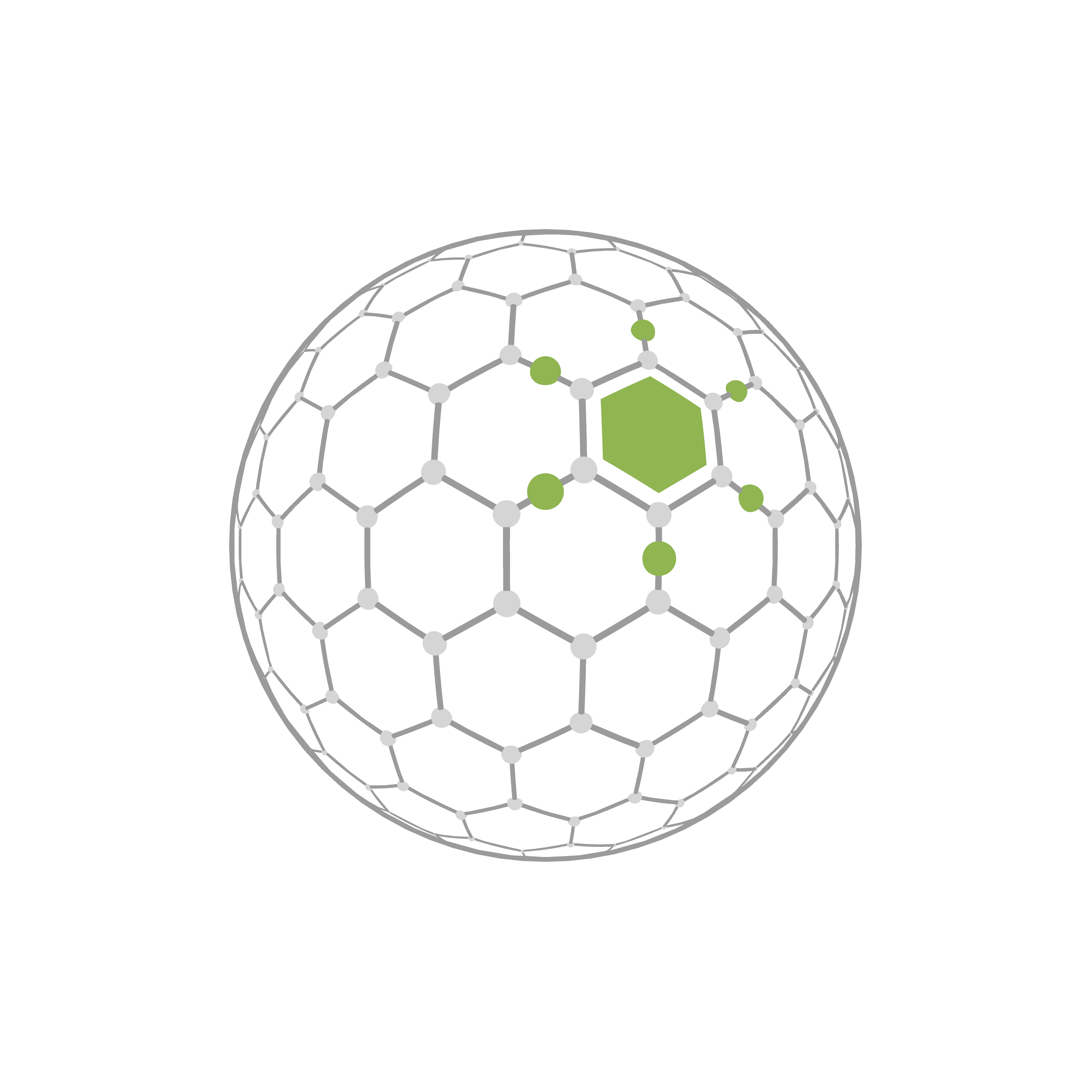}};
          \node[badge=green!60!black, left=0.08in of logo]  (b1) {Artifacts\\Available};
          \node[badge=purple!70!black, left=0.08in of b1]   (b2) {Expert\\Verified};
          \node[badge=green!50!black,  left=0.08in of b2]   (b3) {AI\\Assisted};
          \node[badge=blue!70!black,   left=0.08in of b3]   (b4) {Human\\Conceived};
        \end{tikzpicture}%
      }%
    }%
  }%
  \fi
}

\newcommand{\maketrcover}{%
  \ifarxiv\else
  \thispagestyle{empty}
  \begin{fullwidth}
  \vspace*{2in}
  \begin{center}
    {\Large\sffamily\bfseries D\AE D\AE LUS Technical Report}\\[1.5em]
    {\LARGE\sffamily\bfseries Circumventing the FLP Impossibility Result\\[0.3em]
     with Open Atomic Ethernet}\\[2em]
    {\large \trauthor\,,\;\traffiliation}\\[1em]
    {\normalsize \trversion\quad---\quad\trdate}
  \end{center}

  \vspace{2em}
  \noindent\rule{\linewidth}{0.4pt}

  \vspace{1em}
  \footnotesize
  \begin{description}[leftmargin=1.2in, style=sameline, font=\normalfont\scshape]
    \item[Status:]       Draft --- not yet submitted for review
    \item[Filename:]     \texttt{\trfilename}
    \item[Keywords:]     FLP impossibility, consensus, Open Atomic Ethernet, OAE,
                         Shannon Slots, bilateral swap, FITO, category mistake
    \item[Related:]      OAE Specification (OCP),
                         Leibniz Bridge Synthesis (DAE internal)
    \item[License:]      \textcopyright\ 2026 \trauthor, \traffiliation.
                         All rights reserved.
  \end{description}

  \vspace{1.5em}
  \noindent\rule{\linewidth}{0.4pt}

  \vspace{2em}
  \begin{center}
    \normalsize\itshape
    This cover page may be discarded when printing.\\
    The paper begins on the following page.
  \end{center}

  \end{fullwidth}
  \clearpage
  \fi
}

\fancypagestyle{plain}{%
  \fancyhf{}%
  \fancyfoot[L]{\small\textsc{\trshorttitle}}%
  \fancyfoot[C]{\small\thepage\ of \pageref{LastPage}}%
  \fancyfoot[R]{\scriptsize\texttt{\trfilename}}%
  \ifarxiv\else
  \fancyfoot[L]{\raisebox{-0.8in}{\tiny\texttt{\trversion\ -- \trdate}}}%
  \fancyfoot[R]{\raisebox{-0.8in}{\tiny\texttt{\trfilename}}}%
  \fi
}
\title{Circumventing the FLP Impossibility Result\\with Open Atomic Ethernet}
\author{Paul Borrill, D\AE D\AE LUS}
\date{February 2026}

\begin{document}
\maketrcover
\setcounter{page}{1}%  title page = page 1 (cover page is discardable page 0)
\maketitle
\placebadges
\thispagestyle{plain}

% ── Margin TOC (Leibniz Bridge standard) ──────────────────────────
\marginnote{%
  \hypersetup{linkcolor=linkgreen}%
  \small
  \setcounter{tocdepth}{2}%
  \etocsetstyle{section}{}{\par\vspace{1em}\noindent\leftskip=0em%
    \parbox[c]{1.5em}{\centering\etocnumber}%
    \parbox[c]{\dimexpr\marginparwidth-3em\relax}{\linespread{0.75}\selectfont\raggedright\textbf{\textsc{\etocname}}}%
    \hfill\etocpage\par}{}{}%
  \etocsetstyle{subsection}{}{\par\vspace{0.5em}\noindent\leftskip=1.5em%
    \parbox[c]{1.5em}{\centering\etocnumber}%
    \parbox[c]{\dimexpr\marginparwidth-3em\relax}{\linespread{0.75}\selectfont\raggedright\textsc{\etocname}}%
    \hfill\etocpage\par}{}{}%
  \etocsettocstyle{}{}%
  \tableofcontents
}

\begin{abstract}
The Fischer--Lynch--Paterson (FLP) impossibility result is widely regarded as one of the most fundamental negative results in distributed computing: no deterministic protocol can guarantee consensus in an asynchronous system with even one faulty process. For forty years, the field has treated this as an immovable constraint, designing around it with randomized protocols, failure detectors, and weakened consistency models. This essay argues that FLP is not a law of physics but a theorem about a particular \emph{system model}---and that Open Atomic Ethernet (OAE) circumvents it by rejecting the asynchronous model at its foundation. We introduce the term \emph{bisynchronous} to describe OAE's key property: bounded-time bilateral resolution in which both parties reach common knowledge of outcome at every round boundary---a strictly stronger guarantee than synchrony alone. By constructing a bisynchronous, swap-based protocol at Layer~2, OAE sidesteps the load-bearing assumptions of FLP's asynchronous model, achieving deterministic atomic coordination without violating any impossibility result.
\end{abstract}

% ------------------------------------------------------------------
\FloatBarrier
\section[The FLP Result]{The FLP Result: What It Actually Says}
% ------------------------------------------------------------------

The 1985 paper by Fischer, Lynch, and Paterson\sidecite[2in]{FLP85}{Fischer et~al., \emph{JACM}, 1985} is routinely cited as proof that ``distributed consensus is impossible.'' This characterization, while not technically wrong, obscures a critical detail: the result is a theorem about a specific system model, not a universal physical law.

\begin{theorem}[FLP, 1985]
No deterministic protocol can guarantee consensus among $n \geq 2$ processes in an asynchronous system if even one process may crash.
\end{theorem}
\marginnote[1.5cm]{\footnotesize The FLP result is more precisely: impossibility of \emph{deterministic} consensus in an asynchronous message-passing system tolerating one crash failure.}

The proof proceeds by the \emph{bivalency argument}. Starting from an initial configuration where both decision values (0 and~1) are reachable---a bivalent configuration---the adversary exploits asynchrony to keep the system perpetually undecided. The adversary's power is simple: it controls message scheduling. Because the system is asynchronous, there is no upper bound on message delivery time, and therefore no way for any process to distinguish a slow process from a crashed one. The adversary uses this ambiguity to indefinitely defer any decisive step.

The result rests on three explicit assumptions:

\begin{enumerate}[leftmargin=*, itemsep=3pt]
\item \textbf{Asynchrony:} There is no upper bound on message delivery time or relative process speeds.
\item \textbf{Determinism:} The protocol's state transitions are deterministic functions of current state and received messages.
\item \textbf{Crash failure:} At least one process may halt permanently at any point, without warning.
\end{enumerate}

Each of these is a \emph{modeling choice}, not a fact of nature. The asynchronous model was chosen because it appeared to be the weakest---and therefore most general---set of assumptions under which one might hope to solve consensus. That generality is real, but it is purchased at the cost of assuming away the very structure that makes coordination possible.

\marginnote{\footnotesize Note what FLP does \emph{not} say: it does not say consensus is impossible in synchronous systems. It does not say consensus is impossible with stronger primitives. It does not say coordination is physically impossible.}

% ------------------------------------------------------------------
\FloatBarrier
\section[The Load-Bearing Assumption]{The Assumption That Does the Work}
% ------------------------------------------------------------------

Of the three FLP assumptions, asynchrony is the load-bearing one. The adversary's entire strategy depends on the inability to distinguish slow from dead. If processes could detect crashes within bounded time, the bivalency argument collapses: the adversary can no longer indefinitely defer decisions by hiding behind ambiguous delays.

This observation is well known. Chandra and Toueg\sidecite{CT96}{Chandra \& Toueg, \emph{JACM}, 1996} showed that unreliable failure detectors---oracles that eventually identify crashed processes, though they may make temporary mistakes---are sufficient to solve consensus in asynchronous systems. Dwork, Lynch, and Stockmeyer\sidecite{DLS88}{Dwork et~al., \emph{JACM}, 1988} proved that consensus is solvable in partially synchronous systems where message delays are bounded but the bound is unknown. These results demonstrate that the FLP impossibility is \emph{brittle}: even slight relaxations of the asynchronous model restore solvability.

\marginnote{\footnotesize Partial synchrony models assume bounded delays exist but are unknown. This is weaker than synchrony (where the bound is known) but stronger than asynchrony (where no bound exists).}

The conventional response to FLP has been to preserve the asynchronous model and patch it with additional mechanisms: randomization (Ben-Or\sidecite{BenOr83}{Ben-Or, \emph{PODC}, 1983}), failure detectors (Chandra--Toueg), leader election (Paxos\sidecite{Lamport98}{Lamport, \emph{TOCS}, 1998}, Raft\sidecite{Raft14}{Ongaro \& Ousterhout, \emph{ATC}, 2014}), or weakened consistency guarantees (eventual consistency, CRDTs). Every one of these approaches accepts the asynchronous model as given and attempts to work around its consequences.

OAE takes a different approach entirely. Rather than patching an asynchronous model, it rejects it.

% ------------------------------------------------------------------
\FloatBarrier
\section[The Category Error]{The Category Error in Conventional Networking}
% ------------------------------------------------------------------

The deeper question is: why does conventional networking operate in the asynchronous model at all? The answer is historical, not physical.

Ethernet was designed as a shared-medium broadcast protocol where collisions were expected and frame loss was normal. When switched Ethernet replaced shared media, the point-to-point links between switch and host became deterministic physical channels---but the protocol semantics inherited from the broadcast era were never updated. Layer~2 remained fire-and-forget: frames are transmitted into the void, with no bilateral acknowledgment, no transactional semantics, and no guaranteed delivery.

\marginnote{\footnotesize The ``fire-and-forget'' semantics of Layer~2 Ethernet are a historical artifact of CSMA/CD, not a physical necessity of point-to-point links.}

This creates an \emph{artificial} asynchrony. The physical link between two directly connected endpoints has bounded propagation delay (determined by the speed of light in the medium and the cable length), bounded serialization delay (determined by the line rate and frame size), and deterministic error detection (via CRC). The physics of the link provides bounded propagation and serialization delay. It is only the protocol that imposes asynchrony by choosing not to exploit this structure.

The consequence is that every layer above Layer~2 inherits the asynchronous model by default. TCP must deal with unbounded delays, lost segments, and reordering---not because the physics demands it, but because Layer~2 discards the acknowledgment information that would prevent these pathologies. Higher layers then spend enormous complexity reconstructing, approximately and probabilistically, what the physical layer knew all along.

This is the category error: treating a semantic property (transactional atomicity) as something that can be constructed from syntactic mechanisms (checksums, sequence numbers, retransmission timers) layered over a substrate that has already destroyed the relevant information. It is analogous to attempting to determine the color of an object from a black-and-white photograph. The information existed at the moment of capture but was discarded by the recording medium. No amount of image processing can recover it.

% ------------------------------------------------------------------
\FloatBarrier
\section[OAE's Model]{OAE's Synchronous Model}
% ------------------------------------------------------------------

OAE circumvents FLP by constructing a protocol that violates the asynchronous assumption at its root: the physical link layer. The mechanism has three components.

\subsection[Shannon Slots: Register Reconciliation, Not Time Slots]{Shannon Slots: Register Reconciliation, Not Time Slots}

A critical distinction must be drawn at the outset. Slotted Aloha and Ethernet quantize \emph{time on the medium}. OAE quantizes \emph{ownership of state between endpoints}. The former solves contention; the latter solves ambiguity. These are not incremental differences---they are different problem definitions.

The term ``slot'' in networking history refers to a \emph{time window}---from Aloha to Slotted Aloha to Ethernet, the slot is when a node may attempt to transmit on a shared medium. Metcalfe's original Ethernet was driven by the Aloha insight: discretizing time into slots reduces collisions on a shared channel. That model was correct for its purpose, and it was locally optimal for contention-based MAC on a shared medium.

\ifarxiv\def\metcalfeoffset{-6.5in}\else\def\metcalfeoffset{-4.5in}\fi
\marginnote[\metcalfeoffset]{%
  \centering
  \includegraphics[width=0.9\marginparwidth]{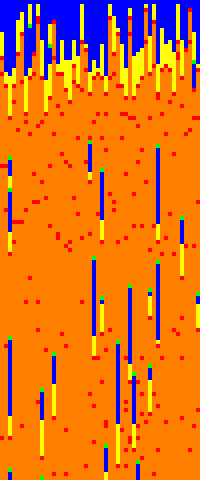}\\[3pt]
  \scriptsize\textit{Aloha retransmission catastrophe} (Metcalfe, 2022\footnote{R.~Metcalfe, ``ArpaAlohaEtherInternet: distributed algorithms for broadcast queues,'' \emph{Wolfram Community}, 2022. \url{https://community.wolfram.com/groups/-/m/t/2576207}}). Blue~=~thinking, orange~=~retransmitting. As load increases, retransmissions dominate and throughput collapses. Metcalfe's collision-domain ontology was not wrong---it persisted beyond its original context.%
}

But when switched Ethernet replaced shared media, the physical reality changed: point-to-point, full-duplex links have no collisions, no contention, no shared channel. Yet Layer~2 semantics inherited from the broadcast era were never updated. The collision-domain ontology---where the wire is the scarce object and time must be partitioned to regulate access---persisted into a world where it no longer applies.

OAE's Shannon Slots are a fundamentally different object. A Shannon Slot is not a time window on a shared medium---it is a \emph{paired register} at each end of a point-to-point link. In OAE, the wire is not the scarce object; \emph{ambiguity} is the scarce object. OAE is a \emph{slot reconciliation protocol}: it atomically reconciles the contents of register pairs across the link. The protocol does not schedule access; it ensures that both sides' registers reach a consistent state at every reconciliation boundary.

The time bound~$\Delta$---the minimum interval for one complete bilateral reconciliation, determined by line rate and cable length---is a \emph{physical consequence} of register reconciliation, not the defining abstraction. What matters is not the duration but the outcome: after each reconciliation, both endpoints know the state of both registers. This gives three properties:

\begin{enumerate}[leftmargin=*, itemsep=3pt]
\item \textbf{Bilateral resolution:} Both registers reach one of two states---$(M, M)$ or $(\emptyset, \emptyset)$---with no ambiguous intermediate ownership.
\item \textbf{Bounded completion:} Reconciliation completes within~$\Delta$, which is known and fixed.
\item \textbf{Informative silence:} If no register update arrives within~$\Delta$, none was sent. Silence is a definitive outcome, not an ambiguity.
\end{enumerate}

This is precisely the model under which Halpern and Moses\sidecite{HalpernMoses90}{Halpern \& Moses, \emph{JACM}, 1990} proved that common knowledge is achievable: a synchronous system with guaranteed delivery within time~$\Delta$, where common knowledge---the infinite regress of ``I know that you know that I know\ldots''---is established at round boundaries. In asynchronous systems it is provably unattainable in finite time; in OAE's register reconciliation model, common knowledge of slot outcome is achieved at each reconciliation boundary.

\subsection[Swap, Not Send: The Universal Primitive]{Swap, Not Send: The Universal Primitive}

The register-reconciliation framing makes the distinction from conventional networking sharp. OAE does not \emph{send} messages. It performs \emph{bilateral register swaps}: atomic exchanges between NIC register pairs at each end of the link.

\begin{figure}[ht]
  \centering
  \includegraphics[width=\textwidth]{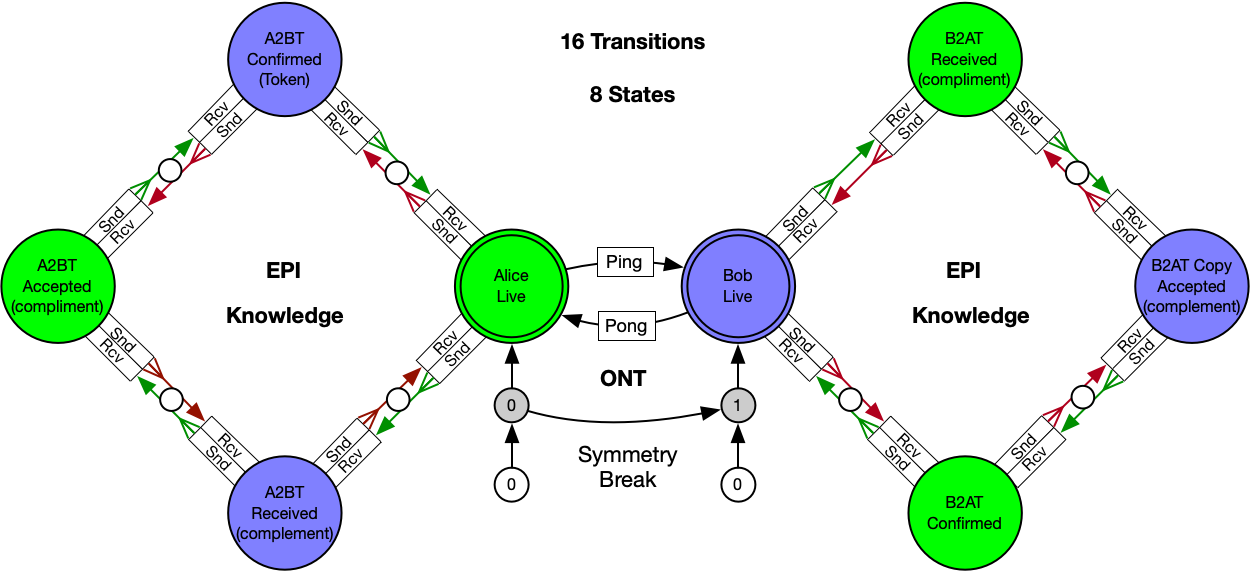}
  \caption{OAE bilateral swap as Petri net. Alice and Bob hold symmetric EPI (epistemic) knowledge states. The Ping/Pong heartbeat drives ONT (ownership) transitions. 16~transitions, 8~states; the symmetry break resolves to a single committed outcome.}
  \label{fig:dual-diamond}
\end{figure}

The Petri net formalism makes the symmetry explicit: transitions are paired and ownership is conserved across reconciliation.

In a conventional ``send'' operation, the sender unilaterally transmits a frame. The frame enters an ambiguous intermediate state---``in the wire''---where ownership is undefined. If the frame is lost, the sender does not know; if it arrives corrupted, the receiver may or may not report this. The asymmetry between sender and receiver creates the information gap that drives the FLP adversary.

In an OAE register swap, both endpoints participate symmetrically. The reconciliation has only two permitted outcomes:

\begin{itemize}[leftmargin=*, itemsep=3pt]
\item $(M, M)$: Both registers hold the message. Transaction committed.
\item $(\emptyset, \emptyset)$: Neither register holds the message. Transaction aborted.
\end{itemize}

No intermediate state is permitted. There is no ``message in flight'' with ambiguous ownership. The register swap is a single atomic event from the perspective of both endpoints.

This connects directly to Herlihy's consensus hierarchy\sidecite{Herlihy91}{Herlihy, \emph{TOPLAS}, 1991}. Read/write registers have consensus number~1---they cannot solve consensus between even two processes. The reason is commutativity: reads and writes to different locations commute, so neither process can detect ordering. Memory-to-memory swap, by contrast, has infinite consensus number. It is a \emph{universal} primitive: it can implement any concurrent object for any number of processes. We treat the bilateral register swap as implementing the shared-memory swap primitive at the link abstraction layer. OAE thus builds its coordination on a universal primitive rather than a consensus-number-1 primitive.

\subsection[Deterministic Error Handling: No Timeouts]{Deterministic Error Handling: No Timeouts}

In the asynchronous model, the standard response to potential failure is the timeout: if no response arrives within~$T$ seconds, assume failure. Timeouts are the mechanism by which asynchronous protocols attempt to simulate synchrony, and they fail at it for a well-known reason: the choice of~$T$ is arbitrary. Too short, and the protocol falsely declares failures; too long, and it stalls unnecessarily. The conflation of ``slow'' with ``dead'' is the operational manifestation of the FLP impossibility.

\marginnote{\footnotesize Timeouts are the operational manifestation of asynchrony. Eliminating them requires eliminating asynchrony itself.}

OAE eliminates timeouts by eliminating their need. Because every Shannon Slot has a known, bounded duration, the absence of a response within~$\Delta$ is not ambiguous---it is a definitive negative outcome. The protocol does not need to guess whether a peer is slow or dead; it knows, with certainty, at every slot boundary.

Combined with credit-based flow control (which prevents buffer overflow and therefore silent frame drops), this creates a system where every message has a deterministic, bounded-time outcome. No retransmission timers. No exponential backoff. No retry storms.

\subsection[Bisynchrony: Beyond the Synchronous/Asynchronous Dichotomy]{Bisynchrony: Beyond the Synchronous/Asynchronous Dichotomy}

The term ``synchronous'' is significantly overloaded in computing. It may refer to clocked logic (synchronous DRAM), blocking API calls (synchronous I/O), bounded-time message delivery (the DLS model), or lockstep execution (BSP). When we say OAE is ``synchronous,'' a software engineer hears ``blocking,'' a hardware engineer hears ``clocked,'' and a distributed systems theorist hears ``bounded-time''---and only the last is close to what we mean.

\marginnote[-2cm]{\footnotesize The three axes that ``synchronous/asynchronous'' incorrectly collapses: (1)~timing model, (2)~API blocking semantics, (3)~epistemic commit symmetry. OAE's innovation is on axis~(3).}

But even ``synchronous'' in the DLS sense is insufficient. A synchronous protocol guarantees bounded delivery but permits \emph{asymmetric knowledge states}: the sender knows it sent, but not whether the receiver committed. The FLP adversary exploits precisely this epistemic gap---the space between what one party knows and what the other knows.

We introduce the term \emph{bisynchronous} to denote a strictly stronger property: bounded-time \emph{bilateral} resolution in which both parties obtain common knowledge of outcome at every round boundary. The prefix \emph{bi-} denotes bilateral epistemic symmetry, not dual-clock alignment. A bisynchronous protocol requires that both endpoints reach the same knowledge state---committed or aborted---at every slot boundary, with no ambiguous intermediate ownership.

\begin{figure}[ht]
  \centering
  \includegraphics[width=\textwidth]{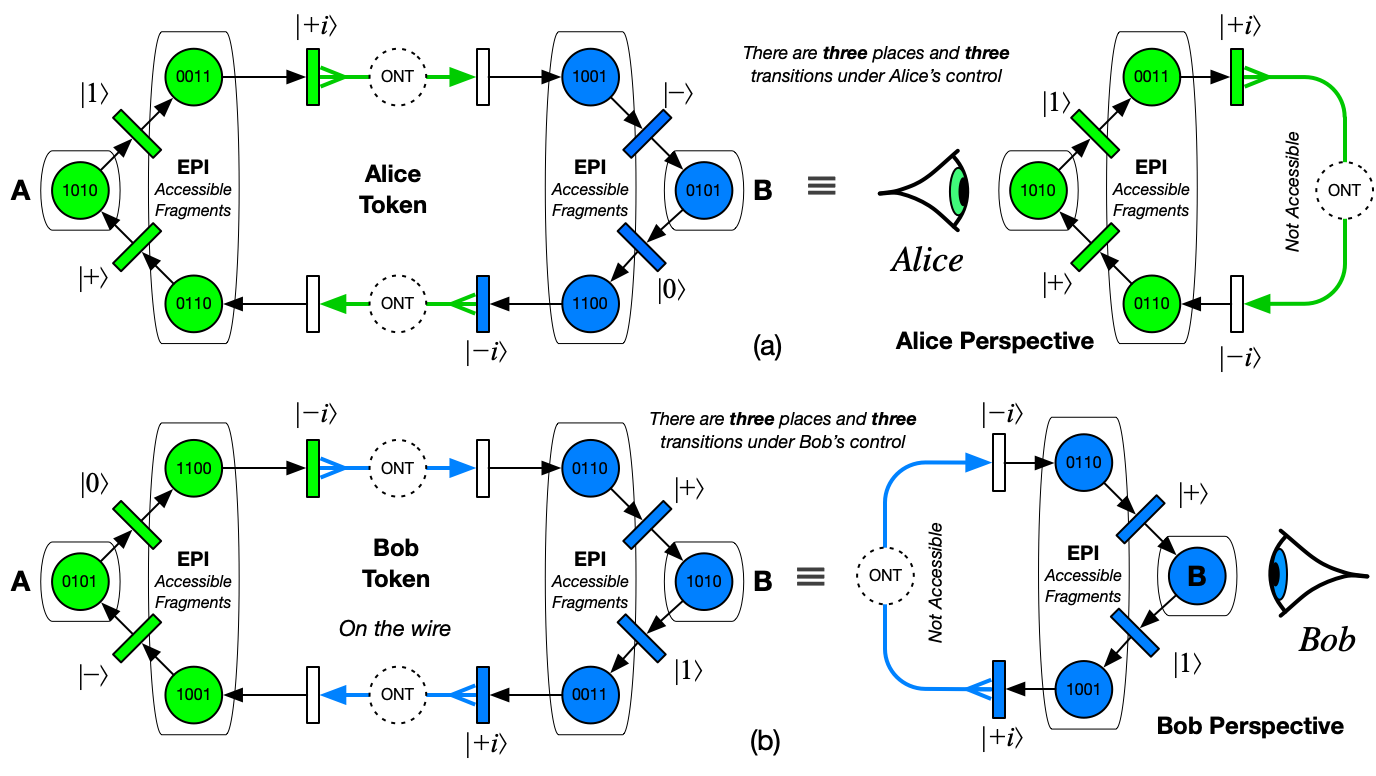}
  \caption{Alice and Bob perspectives of the OAE Petri net. Each party has three places and three transitions under their control. The Spekkens epistemic restriction ensures neither party can distinguish all states---only bisynchronous swap resolves the joint state.}
  \label{fig:spekkens-epi}
\end{figure}

This aligns precisely with the model under which Halpern and Moses~\citep{HalpernMoses90} proved that common knowledge is achievable: a synchronous system with guaranteed delivery within time~$\Delta$. OAE's Shannon Slots instantiate this model at Layer~2, and the bilateral swap ensures that the knowledge gained is symmetric. The result is a three-tier taxonomy:

\begin{description}[leftmargin=*, itemsep=3pt]
\item[Asynchronous:] Unbounded delay, ambiguous silence, no knowledge of peer state.
\item[Synchronous:] Bounded delay, silence is informative, but knowledge may be asymmetric (sender vs.\ receiver).
\item[Bisynchronous:] Bounded delay, silence is informative, and both parties reach common knowledge of outcome at round boundary.
\end{description}

\marginnote[-1cm]{\footnotesize The parallel with Milner's \emph{bisimulation} is structural, not cosmetic. Bisimulation is the equivalence where two systems simulate each other step-for-step; bisynchrony is the communication model where both parties transition simultaneously to a shared knowledge state.}

The FLP adversary's power is precisely the gap between synchronous and bisynchronous: it exploits asymmetric knowledge to keep the system perpetually undecided. Bisynchrony eliminates the asymmetric knowledge gap that the FLP adversary exploits under asynchronous assumptions.

% ------------------------------------------------------------------
\FloatBarrier
\section[Not a Violation]{Why This Does Not Violate FLP}
% ------------------------------------------------------------------

It is important to be precise: OAE does not violate the FLP impossibility result. It circumvents it by operating outside the model to which FLP applies.

FLP proves impossibility in an asynchronous system. OAE is not merely synchronous but \emph{bisynchronous}: the Shannon Slot structure provides bounded-time rounds, and the bilateral swap ensures that both endpoints reach common knowledge of outcome at every slot boundary. The FLP adversary's power derives from exploiting asymmetric knowledge---the gap between what the sender knows and what the receiver knows. Bisynchrony eliminates this gap entirely. The adversary cannot exploit unbounded delays because there are none; it cannot hide crashes behind ambiguous silence because silence within a slot is informative; and it cannot exploit asymmetric knowledge because both parties transition to the same epistemic state simultaneously.

FLP proves impossibility with read/write operations as the coordination primitive. OAE does not use reads and writes. It uses bilateral swaps---a universal primitive with infinite consensus number. The commutativity property that makes reads and writes inadequate for consensus does not apply to swaps.

\begin{table}[ht]
\centering
\small
\begin{tabular}{@{}lll@{}}
\toprule
\textbf{FLP Assumption} & \textbf{Conv.\ Ethernet} & \textbf{OAE} \\
\midrule
Asynchrony      & L2 fire-and-forget  & Shannon Slots \\
R/W primitives  & Unilateral send     & Bilateral swap \\
Crash ambiguity & Timeout detection   & Slot outcomes \\
Message loss    & Silent drops        & Credit flow ctl \\
Knowledge model & Asymmetric          & Bisynchronous \\
\bottomrule
\end{tabular}
\caption{FLP assumptions vs.\ conventional Ethernet and OAE.}
\label{tab:comparison}
\end{table}

The claim is not that FLP is wrong. The claim is that FLP is a theorem about a model that does not apply to OAE. This is not a loophole or a technicality---it is the entire point. The asynchronous model was never a physical necessity; it was an artifact of protocol design choices made in the 1970s. By making different choices at Layer~2, OAE operates in a model where consensus is straightforwardly solvable.

% ------------------------------------------------------------------
\FloatBarrier
\section[Models Are Not Laws]{The Deeper Lesson: Models Are Not Laws}
% ------------------------------------------------------------------

The widespread misinterpretation of FLP as a physical law rather than a model-dependent theorem reflects a broader pattern in distributed systems theory. Results like FLP, the CAP theorem\sidecite{Brewer00}{Brewer, \emph{PODC} Keynote, 2000}\textsuperscript{,}\sidecite{GilbertLynch02}{Gilbert \& Lynch, \emph{SIGACT News}, 2002}, and the Two Generals Problem\sidecite{Gray78}{Gray, \emph{Springer}, 1978} are routinely invoked as if they were laws of nature constraining all possible systems, when in fact they are theorems about specific abstract models.

The CAP theorem, for instance, proves that a distributed system cannot simultaneously guarantee consistency, availability, and partition tolerance. OAE does not claim to defeat CAP---a cable cut still partitions a graph. But OAE changes the engineering regime in which CAP tradeoffs become application-visible, through two mechanisms.

\marginnote{\footnotesize Halpern and Moses~\citep{HalpernMoses90} showed that the Two Generals impossibility---like FLP---depends entirely on the asynchronous model. In a synchronous system, the problem is trivially solvable.}

First, bisynchronous link transactions eliminate ambiguous in-flight states. Every exchange resolves to committed or aborted within one Shannon Slot. A link failure is therefore detected in hundreds of nanoseconds, not inferred after an arbitrary timeout.

Second---and this is the fabric-level consequence of bisynchrony---OAE does not use a Clos network. Clos (fat-tree) fabrics funnel traffic through proprietary switches organized in tiers; a switch failure requires the control plane to reconverge, a process that takes milliseconds to seconds and propagates disruption globally. OAE instead uses an \emph{octavalent mesh}: each cell connects to eight neighbors and every cell forwards packets. Every node is the root of its own spanning tree, named by its private key.

\marginnote{\footnotesize In an $n \times n$ octavalent mesh, the number of spanning trees grows super-exponentially with~$n$ (by Kirchhoff's matrix tree theorem). Even a modest mesh has millions of alternate trees; a $20 \times 20$ mesh has a number beyond comprehension.}

When a link fails, the affected \emph{child cell} heals locally and atomically: it switches to a pre-allocated alternate link toward the root, using only local information. No global routing state is consulted. No control plane reconverges. The healing completes in the next Shannon Slot---hundreds of nanoseconds---far below the timescale at which any higher-layer protocol could detect a partition.

The result is that the dominant class of ``soft'' partitions---local link or cell faults, transient degradations, congestion-induced drops---are detected and healed within bounded time and locality. From the perspective of any protocol above Layer~2, no partition was ever observed. Hard partitions (full graph cuts) remain possible, but the \emph{effective partition surface area} is reduced by orders of magnitude compared to Clos fabrics. This is the networking analog of ECC memory: bit flips occur but are corrected before any software observes them. The memory appears perfect because correction is below the observation threshold.

\smallskip
Bisynchrony is the link primitive; the octavalent mesh is the fabric consequence. Together they reduce the frequency and duration of application-visible partitions---not by preventing physical faults, but by healing them faster than any observer can detect them.

\medskip
The pattern is always the same: an impossibility result is proven in a weak model, the community internalizes it as a universal constraint, and decades of research are devoted to working around it within the model rather than questioning whether the model is appropriate. The possibility that the model itself is a category error---that it captures the wrong abstraction of the underlying physics---is rarely entertained.

Physicists learned this lesson long ago. The Michelson--Morley experiment did not prove that the speed of light is absolute by working within Newtonian mechanics; it prompted the replacement of the Newtonian model with special relativity. The ultraviolet catastrophe was not solved by better calculations within classical thermodynamics; it required quantum mechanics. In each case, the ``impossibility'' was a signal that the model was wrong, not that the phenomenon was forbidden.

OAE applies this principle to networking. The FLP impossibility is not a barrier to be overcome but a signal that the asynchronous model is the wrong abstraction for point-to-point physical links. Replace the model with one that respects the synchronous physics of the link, build on universal primitives rather than consensus-number-1 registers, and the impossibility result no longer applies---not because it was incorrect, but because its assumptions no longer hold.

% ------------------------------------------------------------------
\FloatBarrier
\section[Conclusion]{Conclusion}
% ------------------------------------------------------------------

The FLP impossibility result is a landmark of theoretical computer science, and nothing in this essay disputes its correctness. What OAE challenges is not the theorem but the unexamined assumption that the asynchronous model it describes is the only---or even the appropriate---model for real network links.

By constructing a bisynchronous protocol---Shannon Slots for bounded-time rounds, bilateral swaps for symmetric knowledge, and deterministic slot outcomes that eliminate the ambiguities empowering the FLP adversary---OAE operates in a system model where deterministic consensus is not merely possible but straightforward. At the fabric level, the octavalent mesh with per-node spanning trees ensures that even when physical faults occur, healing is local, atomic, and faster than any higher-layer observer can detect. The protocol does not patch around impossibility; it removes the conditions that create it.

The broader lesson is methodological. Impossibility results are theorems about models, not laws about reality. When an impossibility result blocks progress, the productive response is not always to accept the constraint and engineer around it. Sometimes the productive response is to examine the model and ask whether it captures the right physics. In the case of FLP and conventional Ethernet, the answer is no: the asynchronous model is an artifact of historical protocol choices, not a reflection of the physical link. OAE corrects this by building a bisynchronous protocol layer that respects what the physics has always provided.

% ------------------------------------------------------------------

% Suppress "Float(s) lost" — a cosmetic Tufte-handout issue with dense
% margin content.  The PDF renders correctly; this prevents the error
% from blocking a clean arXiv compile.
\makeatletter
\long\def\@latex@error#1#2{}%
\makeatother

\end{document}